**Parent-offspring Conflict in feral dogs: A Bioassay**

Sreejani Sen Majumder[1], Manabi Paul [1], and Anindita Bhadra[1]

[1] Behaviour and Ecology Lab, Department of Biological Sciences,

Indian Institute of Science Education and Research – Kolkata.

**Address for Correspondence:**

**Anindita Bhadra**

Behavior and Ecology Lab, Department of Biological Sciences,

Indian Institute of Science Education and Research – Kolkata

C.V. Raman Building,

PO BCKV Main Campus, Mohanpur,

PIN 741252, Nadia, West Bengal, INDIA.

Ph: +91 33 25873120 – 207

E-mail: abhadra@iiserkol.ac.in

## Abstract

The parent-offspring conflict theory is an interesting premise for understanding the dynamics of parental care. However, this theory is not easy to test empirically, as exact measures of parental investment in an experimental set-up are difficult to obtain. We have used the Indian feral dog as a model system to test the POC theory in their natural habitat in the context of the mother's tendency to share food given by humans with her pups in the weaning and post-weaning stage. Our behavioural bioassay convincingly demonstrates an increase of conflict and decrease of cooperation by the mother with her offspring over a span of 4-6 weeks. We also demonstrate that the conflict is intentional, and is not influenced by the hunger levels of the pups or the litter size.

## Introduction

Parental care is an indispensible part of development in mammalian species, where mothers suckle their offspring. The offspring often continue to stay with the mother after weaning and the mother continues to share food and shelter with them. It is interesting to note that in the animal world even the most caring of mothers do not extend care towards their offspring for indefinite periods, and at some point of time after weaning the offspring become independent of the mother. The parent-offspring conflict theory delineates a zone of conflict between the mother and her offspring over weaning [1]. We expect that the mother would try to wean her offspring off a little earlier than the offspring would be ready to wean themselves, thereby entering the zone of conflict for a short span of time. Though the theory was originally formulated in the context of weaning, it is also relevant in other contexts where a parent and his/her offspring have conflicting interests. Conflict has been reported over feeding, grooming, travelling, evening nesting, and mating in various species [2-12].

There are several theoretical models that address parent-offspring conflict in different contexts like reproduction, intra-brood competition, sexual conflict, and parental favouritism toward particular offspring [13-16]. Though relatedness between parents, offspring and siblings can be measured easily, it is not easy to measure precisely parental investment and the costs and benefits to the concerned parties in nature. In some studies attempts have been made to quantify parental care in terms of milk ingested by offspring, sometimes as a correlate of weight gain by the individual pups, and sometimes by the duration of suckling [17-21]. However, we have to accept that there is considerable variation in the suckling rates of individual pups, and in hunger levels of individuals, and hence such measures can only provide a rough estimate of parental investment. It is therefore not surprising that empirical tests of the theory in a field set up are sparse, especially in the original context of weaning. The parent-offspring conflict theory has in fact been claimed to be one of the most contentious subjects in behavioural and evolutionary ecology and also one of the most recalcitrant to experimental investigation [22, 23].

Feral dogs are an indispensible part of the urban ecology in many countries. These dogs live in small groups or singly on streets and depend on garbage for their sustenance. Competition over food is quite high, and fights are very common at vats, near roadside food stalls, or when humans occasionally offer a piece of food to the dogs. The feral dogs in India are an excellent study system for addressing various questions in behavioural ecology. These dogs typically breed twice a year, once in the autumn and once in the spring, but a given female usually produces one litter per year. The mother typically spends most of her time with the pups in the early weeks, and only moves out of her shelter to feed for short periods. Her absence increases when the pups gain mobility (Paul and Sen Majumder, unpublished data). Weaning in dogs typically begins when the pups are about 6 weeks old. At this stage the mother begins to refuse to feed the pups while they continue to demand suckling [24]. The feral dog pups begin to eat solid food from around 5 weeks, when the mother gives them regurgitated food and at times hunts small prey to feed her pups. Often mothers with litters are fed by kind people and the pups share this food with their mother when they begin to feed on solids.

We observed that the mother begins to refuse sharing of such food with her pups soon after weaning and the competition between them and the mother seems to increase over the weeks. Since it is extremely difficult to measure parental investment in terms of the actual amount of milk that a pup receives or the energy that the mother spends in caring for her pups, we used the mother's tendency to share food with her pups as a surrogate for parental care. This is especially relevant in these dogs because they are scavengers, and they often beg for food from humans. Competition over food is high, and most of the agonistic interactions within and between dog groups take place at feeding sites [25]. Using the surrogate behaviour of food sharing by the mother for our bioassay, we carried out a field experiment to examine whether post-weaning conflict over food exists between the mother and her pups in the Indian feral dogs.

## Methods

The experiment was performed between January and March 2011 on litters of feral dogs in Kolkata (22° 34' N, 88° 24' E) and at the IISER-K campus at Mohanpur (22° 94' N, 88° 53' E), West Bengal, India. Each litter was observed for a minimum of four and a maximum of six weeks, and only the groups where the mother and at least one pup survived through this period were used for the analysis. Thus we obtained data on 8 out of 14 available litters, varying in size from a single pup to seven pups. We had a total of 8 mothers and 31 pups in our data (Table 1).

The experiment was conducted in two sessions, morning (between 1000-1230h) and evening (between 1530-1700h), on three consecutive days of a week for all litters, thus yielding a total of 210 sessions. The mother and her pups were offered pieces of bread and biscuits (cookies) in the week before the commencement of the experiment and their preference for either were noted. Some groups were choosy about a particular type of food, while others ate whatever was given. We chose to give them either bread or biscuits because they are likely to find these in their day to day foraging at bins and vats, and they also receive these from people quite often when they beg for food. Hence it would be natural for them (or at least the mother) to receive pieces of bread or biscuits from the experimenter, without causing alarm.

The experiment comprised of giving pieces of bread or biscuits to the mother-litter groups, and recording the response of the individuals to the food. Bread was used for only those groups who had showed a clear preference for bread over biscuits. The experimenter offered a piece of food to the group of dogs and waited until it was completely consumed before offering the next piece. The number of offerings made during a session was equal to the number of individuals present in the group at the time of the experiment. The entire experiment was video recorded and the videos were used to tabulate the data at the end of the experiment. For each piece of food offered, we recorded who ate the piece, the latency to first reaction, time taken to eat the piece and the interactions between the mother and pups towards each other. The data was analyzed in StatistiXL version 1.8 and Statistica Release 7.

## Results

The pups were perpetually hungry, but the mothers often chose not to react to the given food. Since there was always at least one pup who displayed an interest in the offered morsel, we used the mother's behaviour towards the pup to define cooperation and conflict. Here we describe six distinct behaviours that the mother showed towards the pups as a response to the giving of food.

Disinterest (DI): The mother did not make an attempt to reach the food in any way or looked away from it.

Allow (AL): The mother looked at the food, but did not move to grab it, allowing the pups to take it.

Offer (OF): The mother took the food and then gave it to the pups, without eating it herself.

Share (SH): The mother took the food and shared it with the pups, and did not show any aggression.

Compete for food (CF): The mother and pups both tried to grab the food and whoever got to the food first took it, without showing any aggression towards the others.

Compete aggressively (CA): The mother barked at or attacked the pups if they tried to get the food, and took the food herself.

Snatch (SN): The mother snatched the food away from the pups and ate it herself.

We pooled all instances of allow, offer and share into the category of co-operation and all instances of compete, compete aggressively and snatch into the category of conflict for our analysis.

(i) Conflict exists:

In all the eight groups we recorded instances of conflict over food between the mother and pups, though the amount of conflict and the time of onset of conflict were quite variable across groups (ESM Figure 1). The pooled data on proportions of conflict and cooperation across the groups revealed that there was an increase in conflict (linear regression; $r = 0.096$, S.E. = 0.026, $p = 0.009$) and decrease in cooperation (linear regression; $r = -0.117$, S.E. = 0.012, $p = 0.000$), with the age of the pups (Figure 1).

(ii) The pups eat more:

We calculated the proportion of food taken by the mother and each of the pups in a group out of the total number of pieces given in a week. In spite of the fact that there was conflict over food between the mother and the pups, when we pooled the data across weeks, the pups were seen to have eaten most of the given food, which was significantly more than that eaten by the mother (Mann-Whitney U test; U = 64.00, df 8,8, p = 0.000). In spite of the cases of conflict over food, the mother always ate less than 50% of the food provided during the experiment in any week. However, when we considered only those cases where there was conflict over a piece of food, the mother took as many pieces as the average pup (Mann Whitney U test; U = 51.0, df = 8,8, p = 0.05) and the pup which took the maximum number of pieces in the litter, designated as the max pup (Mann Whitney U test; U = 51.0, df = 8,8, p = 0.05). The average and max pups in a litter were comparable in terms of the proportion of food taken (Mann Whitney U test; U = 49.5, df = 8,8, p = 0.08) (Figure 2).

(iii) Effect of litter size:

The proportion of food taken by the mother did not depend on the size of her litter (linear regression; r = 0.022, S.E. = 0.03, p = 0.48), which suggests that she took as much of the food as she wanted, irrespective of how many pups competed with her. However, in case of the pups, there was a significant negative correlation between the number of pups in a litter and the proportion of food that the average pup had taken (linear regression; r = -0.103, S.E. = 0.031, p = 0.01). Even in case of the max pup, the proportion of food taken decreased with litter size (linear regression; r = -0.081, S.E. = 0.032, p = 0.04), which suggests that there was considerable competition among the pups (Figure 3). The probability of the mother eating the food to convert it to milk is negligible, as conflict is low in the early weeks of observation, when some suckling occurs, and high when suckling is totally absent (See ESM).

(iv) Conflict decisions are faster:

The latency of first reaction to the food was significantly lower in case of situations of conflict than in cases of cooperation (Mann-Whitney U test; $U = 55.0$, $df = 8,8$, $p = 0.015$) and disinterest (Mann-Whitney U test; $U = 58.0$, $df = 8,8$, $p = 0.005$) for the mother (Figure 4). Thus it appears that in cases of conflict the mother decides to react faster than when she is disinterested or she wants to allow the pups to eat. Thus it is evident that the conflict is quite real, and the mother is actually in competition with the pups over food. The time taken to finish a piece of food whether by a pup or the mother did not vary in the three cases (Mann-Whitney U test; $p > 0.05$).

## Discussion

The cost (in terms of reproductive success) that a mother incurs if she loses an offspring, and the benefit the offspring gets by helping its mother raise its sibling are the two currencies that delineate parent-offspring conflict over parental investment [1]. However, costs and benefits across generations are difficult to measure, and even more difficult to measure is parental investment in terms of the actual amount of milk that a mother produces to feed her offspring or the energy that she spends in caring for them. So, in order to check if such conflict exists in a species, we need to use surrogates for parental investment.

In the Indian feral dogs, competition over food is a reality, and we have observed mothers fight with their offspring over food. Mother dogs lose a lot of weight when they suckle, and they have to compete with other dogs to obtain food in order to regain their energies to support the next litter. We have often observed that people tend to give food to suckling mothers, but such supplies typically do not continue when the pups grow up and begin to fend for themselves. Hence the mother has a short window of time when she can get food without fighting with other adults, and her only competition comes from her own offspring, who are always hungry. Hence we considered food given by humans to be a good surrogate for parental investment in our study system. We expected that the mother should be ready to share food with her offspring at the early stages of weaning, but should compete more often as the pups grow older. So we chose a small window of time, spreading over 4-6 weeks for a litter, beginning around the onset of weaning, when conflict if any, is expected to become evident.

We categorized the behaviour of the mother towards her offspring as a reaction to the giving of the food into cooperation, conflict and disinterest. It was obvious that there is real conflict over food between the mother and her pups as we not only recorded instances of conflict in all the eight groups studied, but also saw that the conflict increased and cooperation decreased over time. It is worthy of note that this increase in conflict and reduction of cooperation was observed in spite of the fact that most of the food given was taken by the pups. This shows in those few cases that the mother actually took the given food she did so by competing with her pups, and not because the pups did not want the food. We also recorded instances of aggression between the mother and her pups over a piece of food. The pups competed among themselves over food, which is evident from the result that the proportion of food obtained by a pup was negatively correlated with its litter size. The fact that the size of the litter did not have any effect on the proportion of food taken by the mother again suggests that she took a piece of food when she wanted it, irrespective of whether the pups were hungry and how many pups were hungry, and her decision was swiftly taken. When she did not take the food, she either allowed her offspring to get the food because she was not interested in it herself, or because she wanted them to have it.

Since the feral dogs depend largely on garbage and on food provided by humans for their sustenance, their diet is not protein rich. Food provided to humans chiefly comprises of carbohydrates in the form of rice, bread and biscuits. Leftovers sometimes contain fish and meat bones, but typically no flesh. Bread and biscuits are the two most common food items that they encounter on a day to day basis. Hence we used these items in our experiment. Since we saw such clear cases of conflict over bread and biscuits, we can expect to see enhanced conflict over more rich food like meat, a situation that would occur in the ancestors of these dogs that probably befriended hunters in the early days of dog domestication. Such conflict would be difficult to observe in pets though, as typically households keep single dogs, and in cases where multiple dogs are kept together, they are usually trained to eat from their designated bowls. The need for conflict might also be redundant in pets, as for them food is not a limiting resource. This study is to the best of our knowledge the first empirical evidence for conflict over given food between the mother and her offspring in any species. The results of this bioassay are exciting because this shows parent-offspring conflict over extended parental investment in a species in its natural habitat, and opens up avenues for many more interesting field experiments using the feral dogs as a model system.

## Acknowledgements

IISER-Kolkata, CSIR India and INSA funded this project. The experiments were carried out equally by SSM and MP. AB designed the experiment, wrote the paper and supervised the work. The work described here complies with the current laws of India. We thank Prof. Raghavendra Gadagkar, Dr. Annagiri Sumana and Dr. Ruchira Sen for their comments on the manuscript.

## References

[1] Trivers R. L. 1964. Parent-offspring conflict. *Am. Zool.* **14**, 249-264.

[2] Duncan P., Harvey P. H., Wells, S. M. 1984. On lactation and associated behaviour in a natural herd of horses. Anim Behav. 32, 255-263. doi: http://dx.doi.org/10.1016/S0003-3472(84)80345-0

[3] Packard J. M., Mech L. D., Ream R. R. 1992. Weaning in an arctic wolf pack: Behavioural mechanisms. *Can. J. Zool.* **70**, 1269-1275.

[4] Bateson P. 1994. The dynamics of parent-offspring relationships in mammals. *Trends. Ecol. Evol.* **9**, 399-403.

[5] Clark C. B. 1977. A preliminary report on weaning among chimpanzees of the Gombe National Park, Tanzania. In *Primate Bio-Social Development* (eds. Chevalier-Skolnikoff S., and Poirier F. E.), pp. 235–260. New York, USA: Garland.

[6] Goodall J. 1968. The behaviour of free-living chimpanzees in the Gombe Stream Reserve. *Anim. Behav. Monogr.* **1,** 161–311.

[7] Horvat J. R., Kraemer H. C. 1982. Behavioral changes during weaning in captive chimpanzees. *Primates* **23**, 488–499. doi: 10.1007/BF02373960

[8] Pusey A.E. 1983. Mother–offspring relationships in chimpanzees after weaning. *Anim. Behav.* **32**, 363–377. doi: http://dx.doi.org/10.1016/S0003-3472(83)80055-4

[9] van de Rijt-Plooij H. H. C., Plooij F. X. 1987. Growing independence, conflict and learning in mother–infant relations in free-ranging chimpanzees. *Behav.* **101**, 1–86. doi: 10.1163/156853987X00378

[10] Yerkes R. M., Tomilin M. I. 1935. Mother–infant relations in chimpanzees. *J. Comp. Psychol.* **20,** 321–348.

[11] Stamps J., Clark A., Arrowood P., Kus B. 1985. Parent-Offspring Conflict in Budgerigars. *Behav.* **94,** 1-40. doi:http://dx.doi.org/10.1163/156853985X00253

[12] van Dijk R. E., Székely T., Komdeur J., Pogány A., Fawcett T. W., Weissing F. J. 2012 Individual variation and the resolution of conflict over parental care in penduline tits. *Proc. R. Soc. B* **279**, 1927-1936. doi:10.1098/rspb.2011.2297.

[13] Parker G. A., Royle N. J., Hartley I. R. 2002. Intrafamilial conflict and parental investment: a synthesis. *Phil. Trans. R. Soc. Lond. B* **357**, 295–307. doi:10.1098/rstb.2001.0950

[14] Macnair M.R., Parker G. A. 1978 Models of parent-offspring conflict. III. Intra-brood conflict. *Anim. Behav.* **27**, 1202-1209. doi:http://dx.doi.org/10.1016/0003-3472(79)90067-8

[15] Mock D. W., Parker G. A. 1997. *The evolution of sibling rivalry*. New York, USA: Oxford University Press.

[16] Lessells C. M. 2002. Parentally biased favouritism: why should parents specialize in caring for different offspring? *Phil. Trans. R. Soc. B* **357**, 381–403. doi:10.1098/rstb.2001.0928.

[17] Riek, A. 2008. Relationship between milk energy intake and growth rate in suckling mammalian young at peak lactation: an updated meta-analysis. *J. Zool.* **274**, 160-170. doi:10.1111/j.1469-7998.2007.00369.x

[18] Pluháček, J.**,** Bartoš, L., Bartošová, J. 2010. Mother-offspring conflict in captive plains zebra (*Equus burchellii*): Suckling bout duration. *Appl. Anim. Behav. Sc.* **122**, s. 127-132. doi: http://dx.doi.org/10.1016/j.applanim.2009.11.009

[19] Gomendio, M. 1991. Parent/offspring conflict and maternal investment in Rhesus macaques. *Anim. Behav.* **42** (6), 993-1005. doi: http://dx.doi.org/10.1016/S0003-3472(05)80152-6

[20] Ahlstrøm, Ø., Wamberg, S. 2000. Milk intake in blue fox (*Alopex lagopus*) and silver fox (*Vulpes vulpes*) in the early suckling period. *Comp. Biochem. Physiol. Part A: Mol. & Int. Physiol.* **127** (2), 225-236. doi: http://dx.doi.org/10.1016/S1095-6433(00)00269-5

[21] Godfray H. C. J., Parker G. A. 1992. Sibling competition, parent-offspring and clutch size. *Anim. Behav.* **43**, 473-490. doi: http://dx.doi.org/10.1016/S0003-3472(05)80106-X

[22] Alexander, R. D. 1974. The evolution of social behaviour. *A. Rev. Ecol. Sys.* **5**, 325-383. doi:10.1146/annurev.es.05.110174.001545

[23] Clutton-Brock. T. H. 1991. *The evolution of parental care.* Princeton University Press, Princeton, NJ.

[24] Malm K., Jensen P. 1997. Weaning and Parent-Offspring conflict in the domestic dog. *Ethol.* **103**, 653-664. doi: 10.1111/j.1439-0310.1997.tb00176.x

[25] Pal S. K., Ghosh B., Roy S. 1998. Agonistic behaviour of free-ranging dogs *Canis familiaris* in relation to season, sex and age. *Appl. Anim. Behav. Sc.* **59**, 331–348. doi: http://dx.doi.org/10.1016/S0168-1591(98)00108-7

| Group Name | Commencement of Experiment | Age of pups (weeks) | Litter size | Female | Male |
|---|---|---|---|---|---|
| Bud 1 | 05.01.2011 | 8 – 13 | 4 | 2 | 2 |
| LEL | 19.01.2011 | 8 – 13 | 2 | 1 | 1 |
| Saltlake 1 | 08.01.2011 | 8 – 12 | 2 | 1 | 1 |
| Saltlake 2 | 15.01.2011 | 8 – 12 | 5 | 3 | 2 |
| BNS | 04.03.2011 | 8 – 12 | 5 | 2 | 3 |
| Canteen | 09.03.2011 | 10 – 15 | 5 | 4 | 1 |
| Bud 2 | 23.02.2011 | 11 – 14 | 1 | 0 | 1 |
| Batanagar | 15.01.2011 | 9 – 14 | 7 | 7 | 0 |
| Total | | | 31 | 20 | 11 |

Table 1: Details of the 8 litters used in the study, with the date of commencement of the experiment.

**Legends to figures**

**Figure 1**

Conflict exists: The proportion of cooperation and conflict over different ages of the pups in weeks. There is an overall increase in conflict and decrease in cooperation between the mother and pups over food.

**Figure 2**

Pups eat more: The mean and standard deviation of the proportion of food taken (of the total number of pieces offered to the group in a week) by the mother, average pup and max pup (the pup who takes the most number of pieces of food) in all the eight groups pooled together.

**Figure 3**

The relationship between the proportion of food taken (of the total number of pieces offered to the group in a week) by the mother (circles), the max pup (triangles) and the average pup (crosses) versus the size of the litter of the respective individuals.

**Figure 4**

Mean and standard deviation of the latency for first reaction to the giving of the food in cases of cooperation, conflict and disinterest by all individuals taken together.

Figure 1

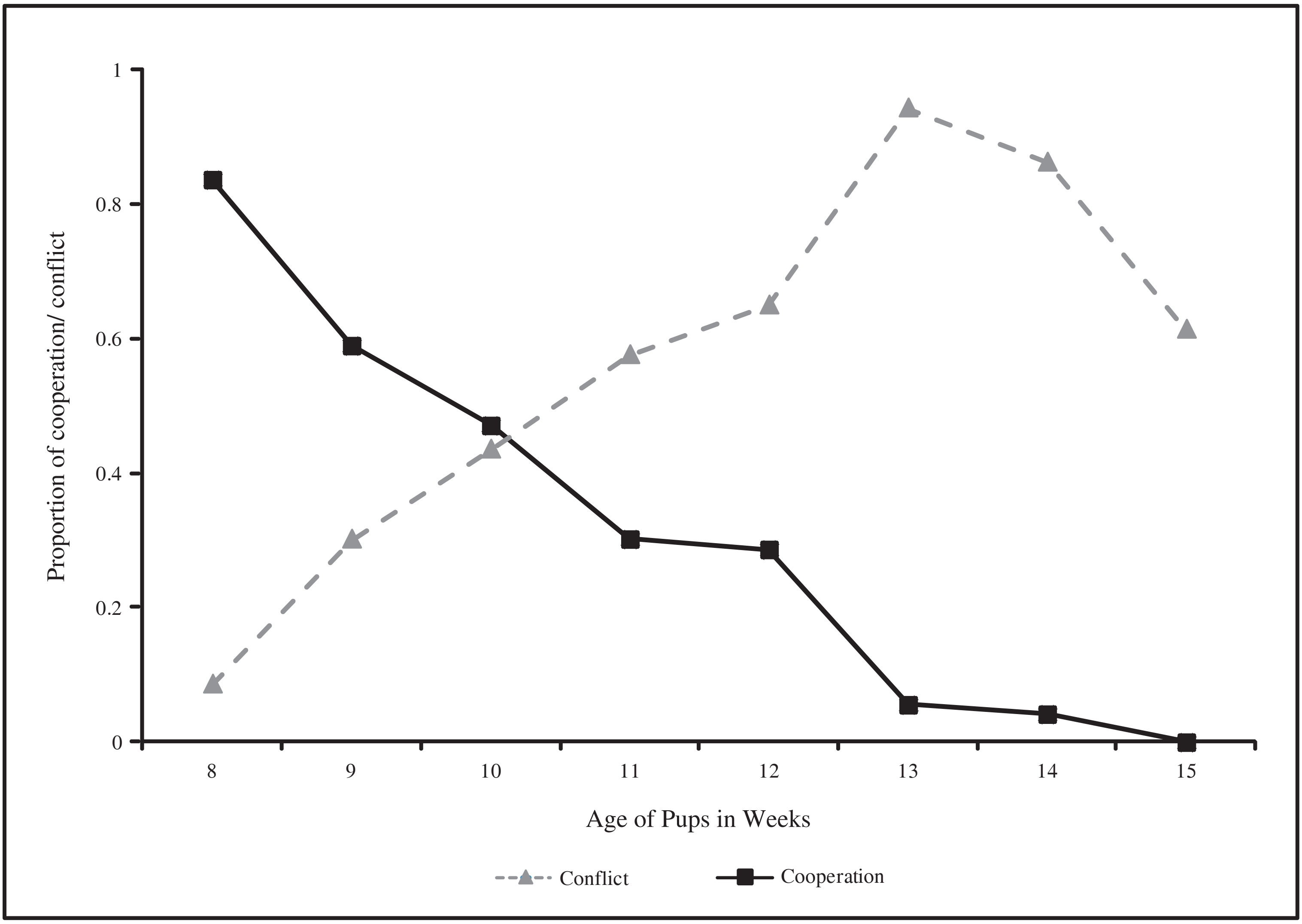

Figure 2

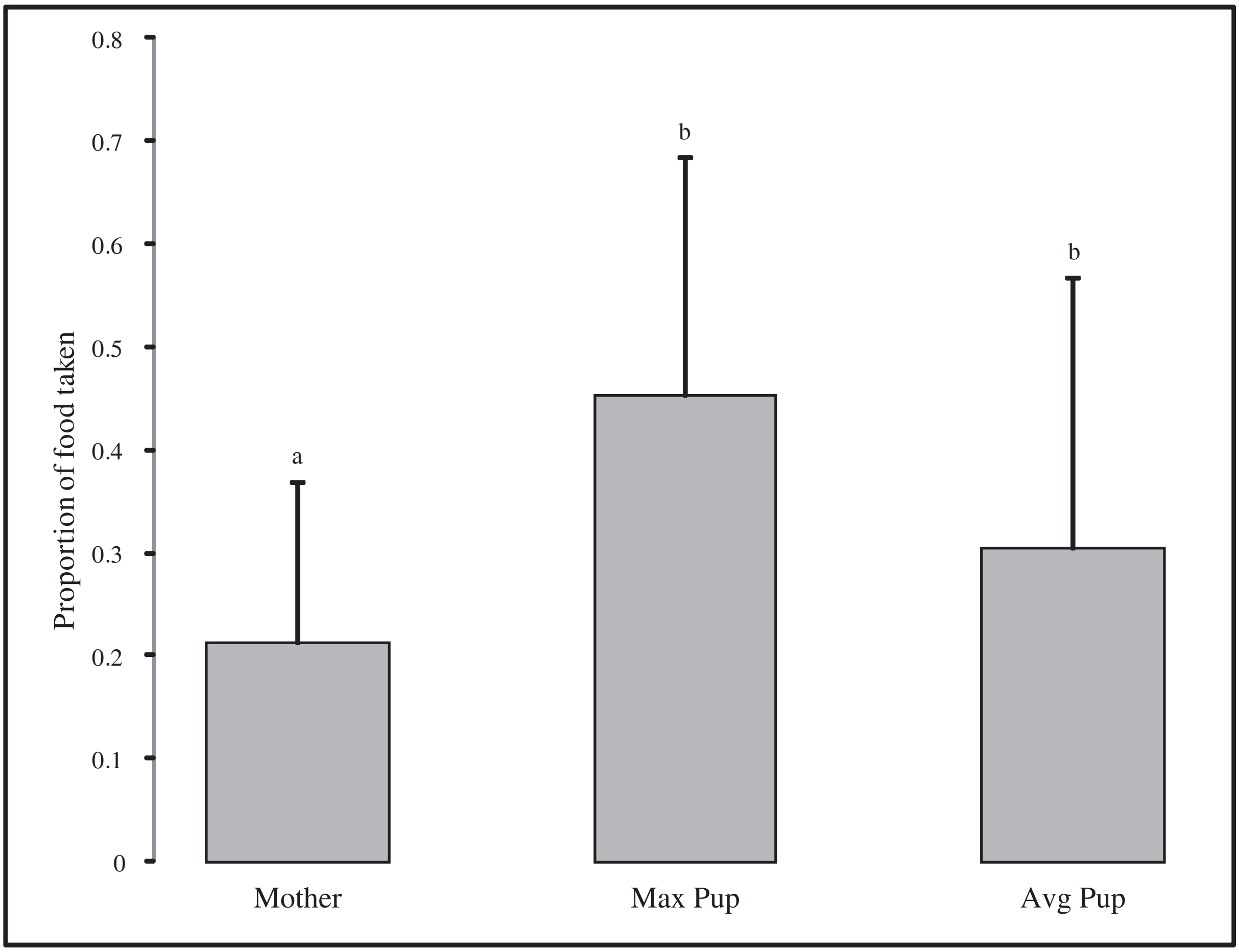

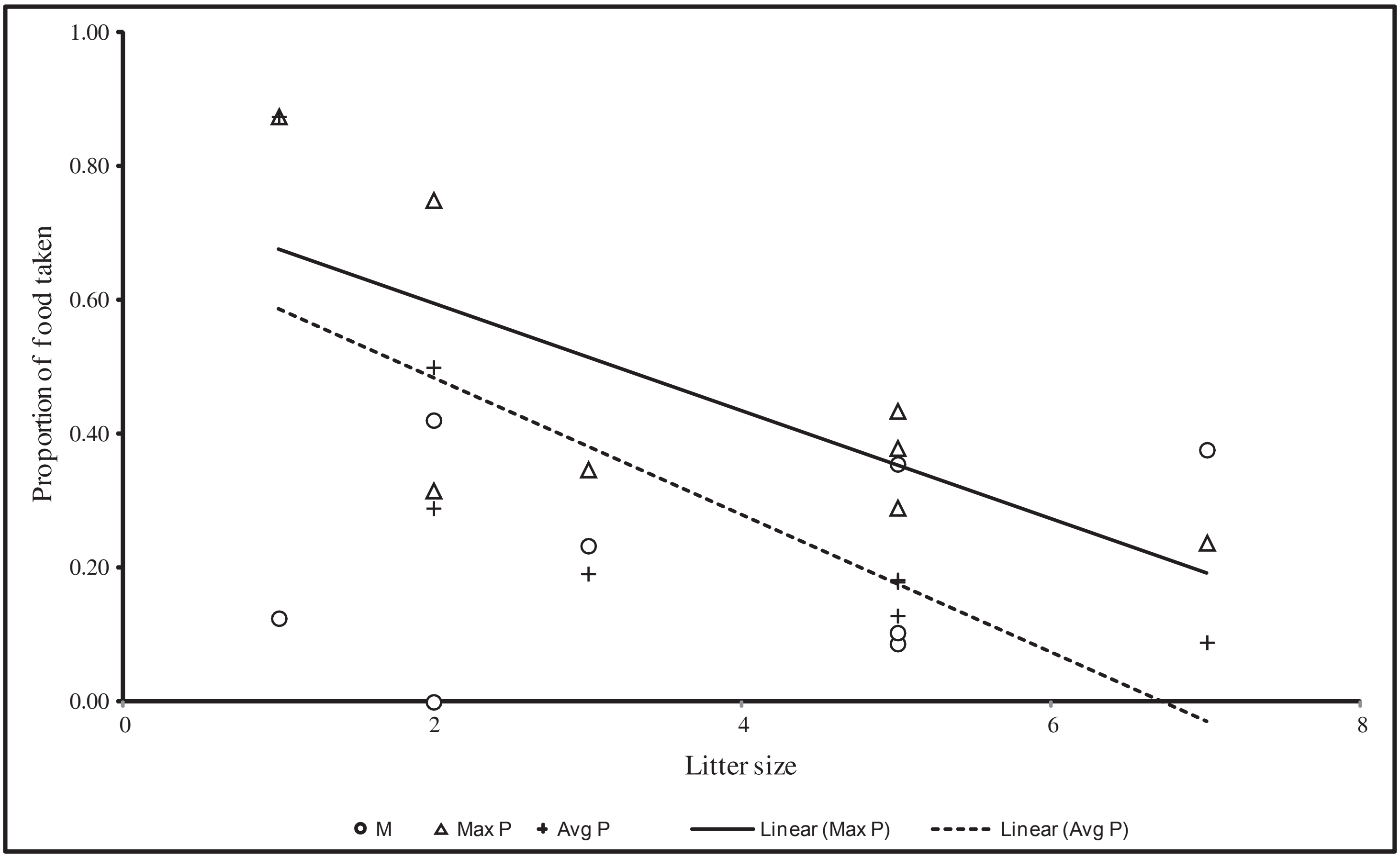

Figure 4

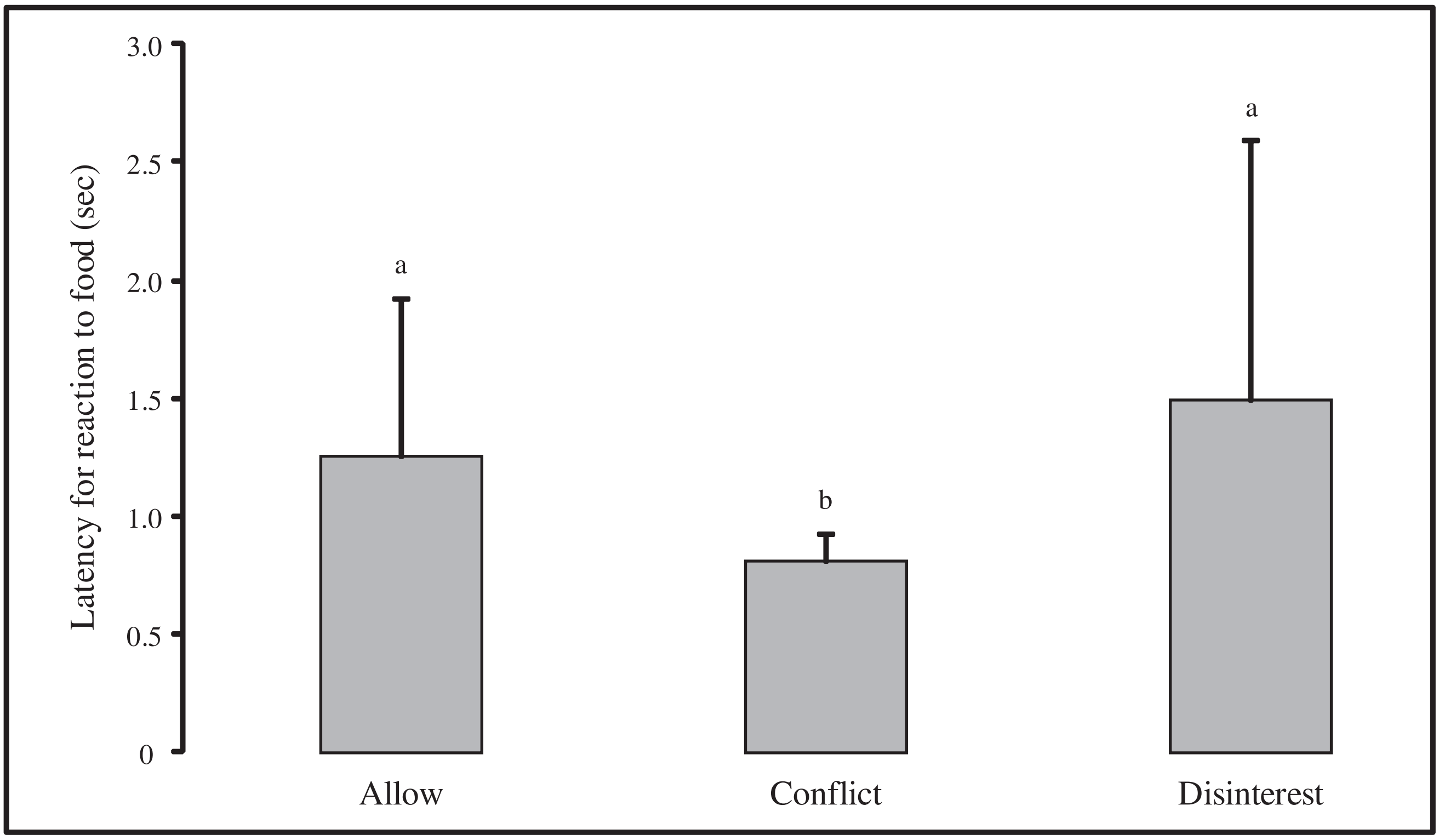

**Parent-offspring Conflict in feral dogs: A Bioassay**

Electronic Supplementary Material

**Conflict exists**

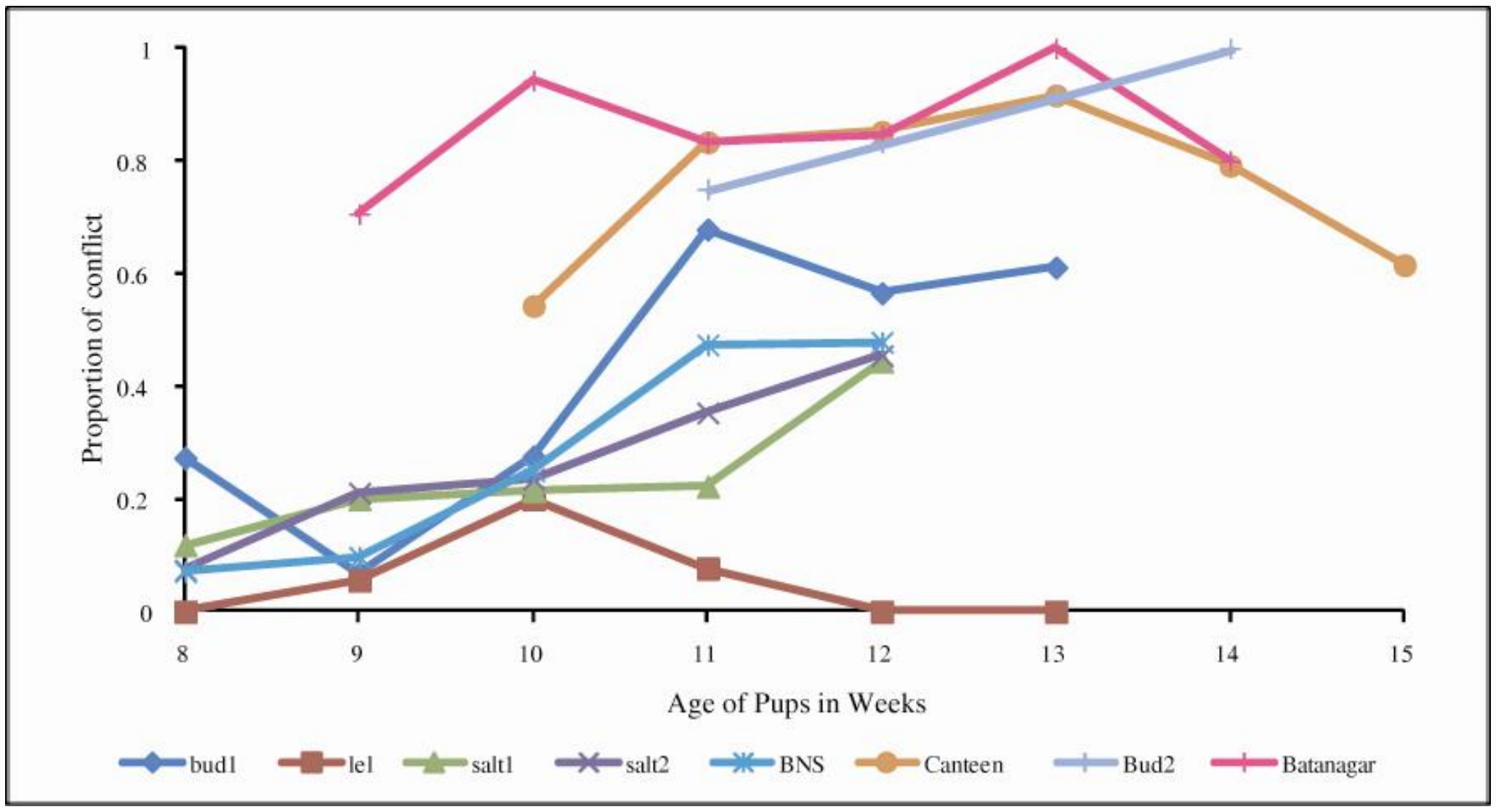

**ESM Figure 1:** The proportion of conflict over different ages of the pups in weeks in the eight groups studied. Conflict is quite variable between the groups, but there is an overall increase in conflict over weeks.

**Suckling**

We counted the number of suckling attempts and refusal to suckle by the mother in all the videos. In the $8^{th}$ and $9^{th}$ weeks there were some suckling attempts, but these attempts decreased drastically over the weeks. Even among the attempts, about half were refused by the mother. The graph shows the mean and SD of all suckling attempts and refusals pooled over litters for each week.

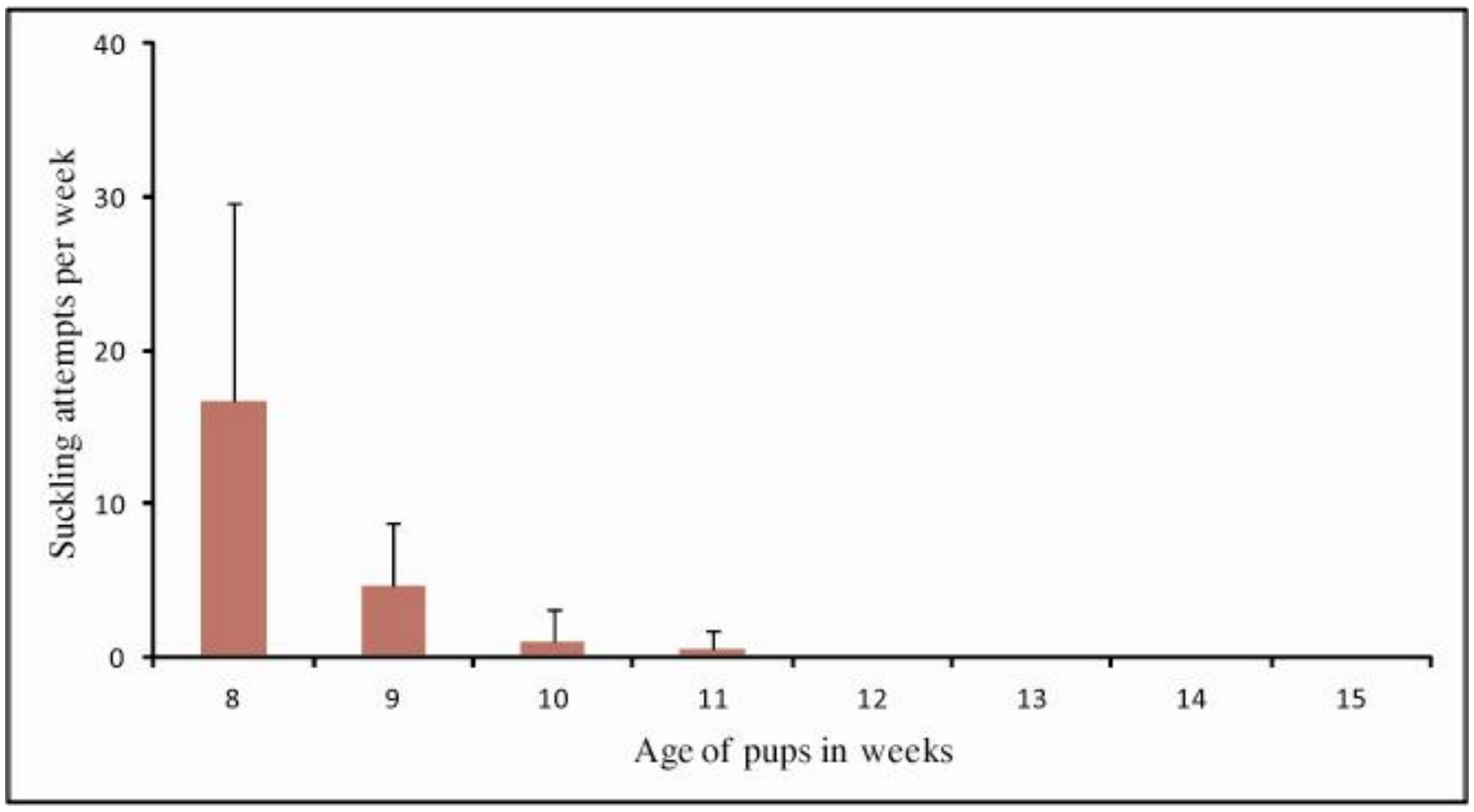

**ESM Figure 2:** Number of suckling attempts by the pups over weeks, pooled across the eight groups. The suckling attempts reduce over weeks and stop at 12 weeks (linear regression, co-eff: 22.421, S.E. = 7.839, $p = 0.029$). It is interesting to note that the maximum conflict was recorded in the 13$^{th}$ week (Figure 1).